\documentclass[conference]{IEEEtran}
\IEEEoverridecommandlockouts 

\usepackage{cite}
\usepackage{amssymb}
\usepackage{stmaryrd}
\usepackage{pifont}

\begin{document}


\title{A Few Remarks About\\Formal Development of Secure Systems}

\author{\IEEEauthorblockN{\'Eric Jaeger}
\IEEEauthorblockA{Direction centrale de la s\'ecurit\'e des syst\`emes d'information\\
51 boulevard de la Tour-Maubourg\\75700 Paris 07 SP, France}
\and
\IEEEauthorblockN{Th\'er\`ese Hardin\thanks{This work is supported in part by the
\emph{Agence Nationale de la Recherche} under grant ANR-06-SETI-016 for the \emph{SSURF}
Project.}}
\IEEEauthorblockA{LIP6, Universit\'e Pierre et Marie Curie (Paris 6)\\
4 place Jussieu\\75252 Paris Cedex 05, France}}

\maketitle

\IEEEpeerreviewmaketitle


\begin{abstract}
Formal methods provide remarkable tools allowing for high levels of confidence in the
correctness of developments. Their use is therefore encouraged, when not required, for the
development of systems in which safety or security is mandatory. But effectively specifying a
secure system or deriving a secure implementation can be tricky. We propose a review of some
classical `gotchas' and other possible sources of concerns with the objective to improve the
confidence in formal developments, or at least to better assess the actual confidence level.
\end{abstract}


\newcommand{\NAT}
{\ensuremath{\mathbb{N}}}

\newcommand{\REL}
{\ensuremath{\mathbb{Z}}}

\newcommand{\BOOL}
{\ensuremath{\mathbb{B}}}

\newcommand{\Cmd}[1]
{\ensuremath{\mathbf{#1}}}

\newcommand{\restrdom}[2]
{{#2} \vartriangleleft {#1}}

\newcommand{\subdom}[2]
{{#2} \vartriangleleft \hspace{-10pt} - \hspace{4pt} {#1}}

\newcommand{\restrimg}[2]
{{#1} \vartriangleright {#2}}

\newcommand{\subimg}[2]
{{#1} \vartriangleright \hspace{-10pt} - \hspace{4pt} {#2}}

\newcommand{\surcharge}[2]
{{#1} < \hspace{-6pt} + \hspace{2pt} {#2}}

\newcommand{\choice}{\ensuremath{\talloblong}}

\newcommand{\guard}{\ensuremath{\Longrightarrow}}

\newcommand{\fid}[1]
{\ensuremath{\textrm{\emph{#1}}}}

\newtheorem{example}{Exemple}[section]
\newtheorem{remark}{Remarque}[section]
\newtheorem{notation}{Notation}[section]


\section{Introduction}\label{introduction}

Formal methods applied to the development of systems or software are very efficient tools that
allow for high levels of assurance in the validity of the results. By defining languages with
clear semantics and by making explicit how to reason on these languages, they provide a
mathematical framework in which it is possible to ensure the correctness of implementations.
Formal guarantees are often unreachable by more classical approaches; for example they are
exhaustive whereas tests cover only a part of the possible executions.

For these reasons, the use of formal methods is encouraged, when not required, by standards
for the development of systems in which safety is mandatory, e.g. \emph{IEC 61508}
\cite{iec61508}. The situation is similar for the development of secure systems: for the
highest levels of assurance the \emph{Common Criteria} (\emph{CC}, \cite{CommonCriteria})
require the use of formal methods to improve confidence in the development, as well as to ease
the independent evaluation process. Indeed, the verification that the delivered product
complies with its specification is expected to rely, at least to some extent, on a
mechanically checked proof of correctness.

One should not however confuse safety with security. They are overlapping but none includes
the other. Safety mostly aims at limiting consequences of random events (dealing with
probabilities) and security at managing malicious actions (dealing with the difficulty of an
attack). In this paper, we discuss a few concerns more specifically related to the formal
development of secure systems. These concerns are illustrated through simple examples
(sometimes involving a \emph{malicious} developer) in \emph{Coq} \cite{coq:1} or in \emph{B}
\cite{abr:1} but most of them are relevant for other deductive formal methods such as
\emph{FoCal} \cite{focal}, \emph{PVS} \cite{pvs:1}, Isabelle/HOL \cite{nip:1}, etc.

\section{Formal methods}\label{formal_methods}

Standard development processes identify several phases such as specification, design,
implementation and verification operations. Different languages can be used for different
phases; beyond programming languages it is frequent to use natural language, automata,
graphical languages, \emph{UML}, etc. The problem of the correctness of a development can then
be seen as a problem of \emph{traceability} between the various descriptions of the system
produced at different phases.

Formal methods considered in this paper also allow for multiple descriptions of a system;
they differ from standard approaches by enforcing the use of languages with explicit and clear
semantics, and by providing a logical framework to reason on them. Ensuring the correctness
then becomes a mathematical analysis of the traceability (or consistency).

\subsection{About formal specification}\label{formal_spec}

At least two descriptions of a system are generally considered in formal methods, a
\emph{formal specification} and an implementation. The specification is often written in a
logical language (e.g. based on predicates) and is ideally declarative, abstract, high-level
and possibly non-deterministic, describing the \emph{what}. On the other hand, the
implementation is imperative, concrete, low-level, and deterministic, describing the
\emph{how}. To emphasise the difference between declarative and imperative approaches,
consider the specification of the integer square root function,
{\small$\sqrt{n}^2\!\leq\!n\!<\!(\sqrt{n}\!+\!1)^2$}, which is deterministic (for any
{\small$n$} there is at most one acceptable value for {\small$\sqrt{n}$}) but is not a
program: how it is computed is left to the developer.

Simply writing a formal specification is already an improvement compared to standard
approaches. Indeed, by using a formal language, ambiguities are resolved. Furthermore, formal
methods provide ways to check, at least partially, the consistency of the specification.

\subsection{About refinement}\label{refinement}

The process of going from a specification to an implementation, while checking the compliance,
is called \emph{refinement}. This concept captures the activity of designing a system; it
encompasses a lot of subtle activities, including choosing concrete representations for
abstract data or producing operational algorithms matching declarative descriptions. From a
logical point of view, a formal specification describes a family of \emph{models} (that is,
intuitively, implementations) and the refinement process consists of choosing one of those
models.

Any formal method defines, implicitly or explicitly, a form of refinement. The definition
generally ensures that the refinement is transitive (allowing for an arbitrary number of
refinement steps in the development process) and monotone (allowing for the decomposition of a
problem into several sub-problems that will be refined independently).

Formal methods do not automatically produce refinements but explain how to check that a
refinement is valid, that is they ensure that very different objects (a logical description
and an operational implementation) are sufficiently `similar'. To allow for this comparison
the refinement is by nature \emph{extensional}: the objects are seen from a functional point
of view, as black boxes whose only inputs and outputs are relevant. If the specification is to
sort values, any sorting algorithm is valid and any two algorithms are considered equal
(undistinguishable). Note that the word \emph{intensional} is often used to refer to properties
that are not extensional; for example, execution time or memory use for a sorting algorithm
can be considered as intensional.

\subsection{About logic}\label{logic}

Behind any formal method, there is logic -- or more accurately, a logic. It is not the intent
of this paper to discuss at length the various types of logic, once pointed out the common
fact that a specification can be inconsistent.

A specification is \emph{inconsistent} if it is self-contradictory -- a trivial example is to
specify {\small$v$} as a natural value equal to both {\small$0$} and {\small$1$}. Such a
specification is also said to be \emph{unsatisfiable}, that is it does not admit a model in
the logical sense. There are three important points about inconsistent specifications:
\begin{itemize}
\item the detection of inconsistency cannot be automated in the general case (the problem of
satisfiability is \emph{undecidable});
\item an inconsistent specification cannot be implemented;
\item an inconsistent specification can prove any property.
\end{itemize}
Tools implementing formal methods considered in this paper do not even try to detect
inconsistencies (even the trivial example of {\small$v\!=\!0\!=\!1$}), due to the
undecidability as well as because the aim of a formal development being an implementation, any
inconsistency is detected sooner or later. The last point results from the fact that for any
proposition {\small$P$} we have {\small$\Cmd{False}\!\Rightarrow\!P$} (using false assumptions
one can prove anything). The consequences of these points are discussed in this paper.

\section{A short presentation of \emph{B} and \emph{Coq}}\label{intro_B_Coq}

\subsection{About \emph{B}}\label{B}

The \emph{B Method} \cite{abr:1} is a formal method widely used by both the academic world and
the industry. Beyond the well-known examples of developments of safety systems (e.g.
\cite{beh:1}), it is also recognised for security developments.

\emph{B} defines a first-order predicate logic completed with elements of set theory, the
\emph{Generalised Substitution Language} (\emph{GSL}) and a methodology of development in
which the notion of refinement is explicit and central. The logic is used to express
preconditions, invariants and to conduct proofs. The \emph{GSL} allows for definitions of
substitutions that can be abstract, declarative and  non-deterministic (that is,
specifications) as well as concrete, imperative and deterministic (that is, programs). The
following example uses the non-deterministic substitution  {\small$\Cmd{ANY}$} (a `magic'
operator finding a value which satisfies a property) to specify the square root of a natural
number {\small$n$}:
\[\small\begin{array}{l}
\Cmd{ANY}\:x\:\Cmd{WHERE}\:x^2\!\leq\!n\!<\!(x\!+\!1)^2\:\Cmd{THEN}\:\sqrt(n)\!:=\!x
\end{array}\]

The notion of refinement is expressed between \emph{machines} (modules combining a
\emph{state} defined by variables, \emph{properties} such as an invariant on the state and
\emph{operations} encoded as substitutions to read or alter the state) and captures the
essence of program correctness w.r.t. their specification as follows: an implementation
refines a specification if the user cannot exhibit a behaviour of the implementation that is
not compliant with what is required by the specification. This concept is incorporated into
the methodology by the automated generation of proof obligations at each refinement step, and
is sustained by mathematical justifications not detailed here.

One of the characteristics of the refinement in \emph{B} is that it is independent of the 
internal representation used by the machines, as illustrated by the following example of a
system returning the maximum of a set of stored natural values:
\[\small\begin{array}{l}
\Cmd{MACHINE}\:M_A\\
\Cmd{VARIABLES}\:S\\
\Cmd{INVARIANT}\:S\!\subseteq\!\NAT\\
\Cmd{INITIALISATION}\:S\!:=\!\emptyset\\
\Cmd{OPERATIONS}\\
\quad
store(n)\triangleq\Cmd{PRE}\:n\!\in\!\NAT\:\Cmd{THEN}\:S\!:=\!S\!\cup\!\{n\}\\
\quad
m\!\gets\!get\triangleq\Cmd{PRE}\:S\!\not=\!\emptyset\:\Cmd{THEN}\:m\!:=\!\Cmd{max}(S)\\
\Cmd{END}
\\\\
\Cmd{MACHINE}\:M_C\:\Cmd{REFINES}\:M_A\\
\Cmd{VARIABLES}\:s\\
\Cmd{INVARIANT}\:s\!=\!\Cmd{max}(S\!\cup\!\{0\})\\
\Cmd{INITIALISATION}\:s\!:=\!0\\
\Cmd{OPERATIONS}\\
\quad
store(n)\triangleq\Cmd{IF}\:s\!<\!n\:\Cmd{THEN}\:s\!:=\!n\\
\quad
m\!\gets\!get\triangleq m\!:=\!s\\
\Cmd{END}
\end{array}\]
The state of the machines is described in the {\small$\Cmd{VARIABLES}$} clause; for the
specification {\small$M_A$} it is a set of natural numbers and for the implementation
{\small$M_C$} a natural number. The {\small$\Cmd{INVARIANT}$} clause defines a constraint over
the state; for {\small$M_A$} it indicates that {\small$S$} is a subset of {\small$\NAT$},
whereas for {\small$M_C$} it describes the \emph{glue} between the states of {\small$M_A$} and
{\small$M_C$} (intuitively claiming that if both machines are used in parallel then
{\small$s$} is always equal to {\small$\Cmd{max}(S)$}). The {\small$\Cmd{INITIALISATION}$}
clause sets the initial state, while the {\small$\Cmd{OPERATIONS}$} clause details the
operations used to read or alter the state. The two machines differ yet {\small$M_C$} refines
{\small$M_A$}: roughly speaking one cannot exhibit a property of {\small$M_C$} which
contradicts one of {\small$M_A$}.

Note the use of the {\small$\Cmd{PRE}$} substitution defining a precondition, that is a
condition that the user has to check before calling an operation. This is an \emph{offensive}
approach; an operation (should not but) can be used when this condition is not satisfied, yet
in such a case there is no guarantee about the result (it may even cause a crash). By
opposition the \emph{defensive} approach is represented in \emph{B} by using \emph{guards}
(that is an {\small$\Cmd{IF}$}) that prevent unauthorised uses. These notions are standard in
formal methods and will be discussed further later in this paper.

\subsection{About \emph{Coq}}\label{Coq}

\emph{Coq} is a proof assistant based on a type theory. It offers a higher-order logical
framework that allows for the construction and verification of proofs, as well as the
development and analysis of functional programs in an \emph{ML}-like language with
pattern-matching.

\emph{Coq} implements the \emph{Calculus of Inductive Constructions}
\cite{DBLP:conf/colog/CoquandP88} and it is frequent in developments to use inductive
definitions. For example, $\NAT$ is defined in the \emph{Peano} style as follows:
\[\small\begin{array}{l}
\Cmd{Inductive}\:\NAT\!:=\!0\!:\!\NAT\:|\:S\!:\!\NAT\!\to\!\NAT
\end{array}\]
This definition means that {\small$\NAT$}  is the smallest set of terms closed under (finite
number of) applications of the \emph{constructors} {\small$0$} and {\small$S$}. {\small$\NAT$}
is thus made of the terms {\small$0$} and {\small$S^n(0)$} for any finite {\small$n$}; being
well-founded, structural induction on {\small$\NAT$} is possible (the induction principle is
automatically derived by \emph{Coq} after the definition of {\small$\NAT$}). The definition
also means that {\small$\NAT$} contains no other values (\emph{surjectivity}) and that
{\small$\forall\:(n\!:\!\NAT),\:0\not\!=\!S(n)$} and
{\small$\forall\:(m\:n\!:\!\NAT),\:S(m)\!=\!S(n)\!\Rightarrow\!m\!=\!n$} (\emph{injectivity}).

Contrary to \emph{B}, there is no enforced development methodology in \emph{Coq}, nor any
explicit refinement process. The user can choose between several styles of specification and
implementation, and has to decide on its own about the properties to be checked. For example
the \emph{weak specification} style consists of defining functions as programs in the
internal \emph{ML}-like language and later checking properties of these functions, as
illustrated here by the division by {\small$2$}:
\[\small\begin{array}{l}
\Cmd{Fixpoint}\:\fid{div2}(x\!:\!\NAT)\!:\!\NAT:=\\
\quad\Cmd{match}\:x\:with\:S(S(x')) \to S(\fid{div2}(x'))\:|\:\_ \to 0\:\Cmd{end}.\\\\
\Cmd{Theorem}\:\fid{div2\_def}\!:\\
\quad forall\:(x\!:\!\NAT),\:n\!=\!2\!*\!\fid{div2}(n)\lor n\!=\!2\!*\!\fid{div2}(n)\!+\!1.\\
\Cmd{Proof.}\\
\quad\ldots\\
\Cmd{Qed.}
\end{array}\]
{\small$\fid{div2}$} is a recursive program (using
{\small$\fid{div2}(x\!+\!2)\!=\!\fid{div2}(x)\!+\!1$}) and
{\small$\fid{div2\_def}$} a property claimed about it; the proof, not detailed here,
ensures that {\small$\fid{div2}$} indeed satisfies {\small$\fid{div2\_def}$}.

\section{Specifying secure systems}\label{spec_security}

We now begin our discussion about developing secure systems using formal methods by
considering more specifically formal specifications of secure systems.

To start with trivial considerations, we first have to note that formal methods offer tools to
express specifications but that there is no way to force a developer to describe the
properties required of the system under development. Clearly, using even the most efficient
formal method without adopting the `formal spirit' is meaningless, as there is no benefit
compared to standard approaches if the formal specification is empty. Note also that a formal
development is a development, and so can also benefit from standard practices such as naming
conventions, modularity, documentation, etc. In the case of formal methods, in fact, the very
process of deriving a formal specification from the book of specifications should be
documented, justifying the formalisation choices and identifying, if any, aspects of the
system left out (as it is generally not reasonable or even feasible to aim at a full
formalisation of a complete system).

Assuming a developer that has adopted the formal spirit, there are further points to care about
in order to develop an `adequate' formal specification for a secure system, that is a
specification not only expressing the required properties, but also ensuring that those
properties are enforced at all stages of the development as well as in any (reasonable)
scenario of usage of the implementation.

Some of the concerns that will be discussed below are applicable for safety or any high
assurance system; for others a malicious developer will be assumed (a threat generally
irrelevant for safety but applicable in security). The ultimate objective of such a malicious
developer is to exploit any weakness of a specification, in order to trap a system while
delivering a mechanically checked proof of compliance. One could consider that such traps
would be detected through code review or testing. Yet, beyond the fact that formal methods are
expected to reduce the need for such activities, we warn the reader that our illustrations are
voluntarily simplistic, and that real life examples of \emph{Trojan Horse} are difficult to
detect.

\subsection{About invalid specifications}\label{invalid_spec}

As pointed out in Par. \ref{logic}, inconsistent specifications are disastrous. Indeed,
whereas inconsistency cannot be automatically detected, it also permits to discharge any proof
obligation expressed -- that is an inconsistent specification can in practice make the
developer life more comfortable. An inconsistent specification is therefore dangerous for
safety developments if a distracted developer fails to notice that its proofs are a little too
easy to produce, and more so for security developments as a malicious developer identifying
such a flaw would be able to prove whatever he wants.

Of course, an inconsistent specification is not implementable. It is therefore possible to
check the consistency by providing an implementation -- any one will do the trick, so even a
dummy implementation is sufficient. Yet there are in security situations in which a formal
specification is mandatory while a formal implementation is not. This is the case for the
\emph{CC}, at some assurance levels, that just require a formal specification of the
\emph{Security Policy}. An undetected inconsistent specification is therefore a possibility.

In \emph{B} the consistency of a specification is partially checked through proof obligations
to be discharged by the developer. Yet the obligations related to the existence of values
satisfying the expressed constraints for parameters, variables and constants are deferred.
Both following specifications are inconsistent, yet all \emph{explicit} proofs obligations can
be discharged (that is, most \emph{B} tools will report a `100\% proven' status):
\[\small\begin{array}{l}
\Cmd{MACHINE}\:\fid{absurd}\_\fid{var}\\
\Cmd{VARIABLES}\:v\\
\Cmd{INVARIANT}\\
\quad v\!\in\!\NAT\:\land\\
\quad v\!=\!0\land v\!=\!1\\
\Cmd{ASSERTION}\:0\!=\!1
\end{array}
\quad\quad
\begin{array}{l}
\Cmd{MACHINE}\:\fid{absurd}\_\fid{cst}\\
\Cmd{CONSTANTS}\:f\\
\Cmd{PROPERTIES}\\
\quad f\!\in\!\NAT\!\to\!\NAT\land\\
\quad \forall\:x,y,\:x\!<\!y\!\Rightarrow\!f(x)\!>\!f(y)\\
\Cmd{ASSERTION}\:0\!=\!1
\end{array}\]
Of course, delaying such proof obligations is justified, as implementing the specification
will force the developer to exhibit a \emph{witness} for {\small$v$} that meets the
specification (a constructive proof that the specification is satisfiable). Therefore,
\emph{B} ensures that any inconsistency is detected, at the latest, at the implementation
stage. But we would like to remind the reader that a formally derived implementation is not
always required. In such a case, one should consider additional manual verifications to check
the existence of valid values for parameters, constants and variables.

Inconsistencies can be rather easy to introduce, accidentally or not, by contradicting
\emph{implicit} hypotheses associated to the used formal method. In \emph{B} for example there
is a clause {\small$\Cmd{SETS}$} that allows for the declaration of abstract sets used in a
machine; one can easily forget that such a set is always in \emph{B} finite and non-empty. If
the developer contradicts one of these implicit hypotheses the specification becomes
inconsistent without any warning by the tool; in fact the automated prover will very
efficiently detect the contradiction as a lemma usable to discharge any proof obligation.
Contradiction of implicit principles of the underlying logic can also be illustrated in
\emph{Coq} with two very simple examples. The first one is a naive tentative of specifying
{\small$\REL$} using {\small$\NAT$}:
\[\small\begin{array}{l}
\Cmd{Inductive}\:\REL:\Cmd{Set}\!:=\!
\fid{plus}\!:\!\NAT\to\REL\:|\:\fid{minus}\!:\!\NAT\to\REL.\\
\Cmd{Hypothesis}\:\fid{zero\_unsigned}\!:\!\fid{plus}(0)\!=\!\fid{minus}(0).
\end{array}\]
Unfortunately, as pointed out in Par. \ref{Coq}, the definition of {\small$\REL$} is not a
specification but an implementation ({\small$\REL$} is the set of all terms of the form
{\small$\fid{plus}(n)$} or {\small$\fid{minus}(n)$}). {\small$\fid{zero\_unsigned}$}
introduces an inconsistency because it contradicts the injectivity principle for the
constructors: for any natural values {\small$n$} and {\small$m$} it is possible to prove in
\emph{Coq} that {\small$\fid{plus}(n)\!\not=\!\fid{minus}(m)$}.

The second example is related to the unexpected consequences of using possibly empty types.
This is illustrated by the following (missed) attempt to define bi-colored lists of natural
values, that is lists with each element marked red or blue:
\[\small\begin{array}{l}
\Cmd{Inductive}\:\fid{blst}\!:\!\Cmd{Set}\!:=\!\fid{red}\!:\!\fid{blst}\to\NAT\to\fid{blst}\\
\quad\quad\quad\quad\quad\quad\quad\quad\quad\:\:
|\:\fid{blue}\!:\!\fid{blst}\to\NAT\to\fid{blst}.
\end{array}\]
In the absence of an atomic constructor for the empty list, {\small$\fid{blst}$} which is the
smallest set of terms stable by application of the constructors is indeed empty. Therefore,
assuming the existence of such a list is inconsistent, and any theorem of the form
{\small$\forall\:(b\!:\!\fid{blst}),\:P$} is provable -- hardly a problem from the developer's
point of view, as he generally tries to prove only those properties he expects. It would be
prudent for any type {\small$T$} introduced in \emph{Coq}, to ensure that it is not empty e.g.
by proving a theorem of the form {\small$\exists\:(t\!:\!T),\:\Cmd{True}$}.

One could also investigate the satisfiability of the preconditions or guards, as defined in
Par. \ref{B}, associated to functions or operations. Indeed, while unsatisfiable preconditions
are not inconsistent, they often represent a form of deadlock, as they mean that it is never
possible to use an operation. They may however be difficult to detect -- there is a famous
example of the database of individuals developed in \cite{abr:1}, in which it is impossible to
insert new entries, as pointed out in \cite{mus:1}, due to the fact that any new individual
introduced in the database should have a father and a mother, while the initial state is an
empty database. To avoid such difficulties the use of adequate tools (animation of models,
model-checking, automatic tests generator, cf.
\cite{CarlierD2008,B07-JaffuelL,ClarkeGL96,PlaggeL07,YuML99}) can be of considerable help.

We would also like to draw the attention of the readers to other types of problematic 
specifications. For example in some cases it may happen that a specification mixes predicates
of the form {\small$P\!\Rightarrow\!Q$} and {\small$P\!\Rightarrow\!\lnot Q$}. Such a
specification is consistent but only as long as {\small$P$} is false; to the least this type
of specification should be considered inappropriate. This is one of the cases for which
specification engineering tools would be considered useful. Such tools associate for example
to a specification {\small$\forall\:x,\:P\!\Rightarrow\!Q$} an additional proof obligation
{\small$\exists\:x,\:P$}; indeed the specification can be \emph{vacuously} true if {\small$P$}
is always false, but it is unlikely that such a specification convey the intended meaning
\cite{DBLP:conf/lpar/SamerV07}.

\subsection{About (mis)understandings}\label{misunderstandings}

Consequences of invalid specifications have been identified and justify establishing
procedures to check consistency. We now discuss the problem of insufficient specifications,
which is more tricky to detect as it generally refers to a difference between a specification
and its intended meaning.

Our very first concern is related to the understanding of the chosen formal method. It is
not reasonable to expect all `users' of formal methods to be expert. One may consider for
example a situation in which a customer convinced by the interest of formal methods may
however not have any in-depth knowledge about any of them. In fact, we would also argue that
should formal methods be more widely used -- definitely something we expect for the future --
they should be accessible to people having received a dedicated training but which are not
expert (this is one of the main objectives of the \emph{FoCal} project
\cite{focal,TFTP04,TPHOL2003,calc03}). The minimum, however, is to ensure that \emph{any} user
has a basic understanding of some of the underlying principles to avoid misinterpretation.

For example, consider the concept of refinement as introduced in Par. \ref{refinement}. The
essence of this concept is to allow to check that specifications and implementations are
`similar'. This similarity should not be too strong, as a refinement relation reduced to
intensional equality of programs (that is, the same code) would be useless. It is for example
standard to consider that computations and transient states are irrelevant. In \emph{Coq} this
is translated by the fact that the equality is modulo {\small$\beta$}-reduction (in other
words, {\small$\fid{square}(3)\!=\!9$} because computing {\small$\fid{square}(3)$} yields
{\small$9$}). Our concern is illustrated in \emph{B} by the following specification of an
airlock system:
\[\small\begin{array}{l}
\Cmd{MACHINE}\:\fid{Sas}\\
\Cmd{VARIABLES}\:\fid{door}_1,\fid{door}_2\\
\Cmd{INVARIANT}\:\fid{door}_1,\fid{door}_2\!\in\!\{\fid{open},\fid{locked}\}\land\\
\quad\quad\quad\quad\quad\quad\quad\:
 \lnot(\fid{door}_1\!=\!\fid{open}\land \fid{door}_2\!=\!\fid{open})\\
\Cmd{OPERATIONS}\\
\quad \fid{open}_1\triangleq \Cmd{IF}\:\fid{door}_2\!=\!\fid{locked}\:\Cmd{THEN}\:
  \fid{door}_1\!:=\!\fid{open}\\
\quad \fid{close}_1\triangleq \fid{door}_1\!:=\!\fid{locked}\\
\quad \fid{open}_2\triangleq \Cmd{IF}\:\fid{door}_1\!=\!\fid{locked}\:\Cmd{THEN}\:
  \fid{door}_2\!:=\!\fid{open}\\
\quad \fid{close}_2\triangleq \fid{door}_2\!:=\!\fid{locked}\\
\end{array}\]
If the underlying principles of the \emph{B} are not understood, one can easily consider that
the {\small$\Cmd{INVARIANT}$} clause in a proven \emph{B} machine is `always true'. Therefore,
any compliant implementation of this specification would be considered safe. Of course, this
is not the case, as we may for example refine the operation {\small$\fid{open}_1$} as follows:
\[\small\begin{array}{l}
\fid{open}_1\triangleq \Cmd{IF}\:\fid{door}_2\!=\!\fid{locked}\:\Cmd{THEN}\\
\quad\quad\quad\quad\quad\fid{door}_1\!:=\!\fid{open};\\
\quad\quad\quad\quad\quad\Cmd{IF}\:\fid{attack}\:\Cmd{THEN}\:
\fid{door}_2\!:=\!\fid{open};\fid{wait};\fid{door}_2\!:=\!\fid{locked}
\end{array}\]
where {\small$\fid{wait}$} is a passive but slow operation and {\small$\fid{attack}$} any
condition the malicious developer can imagine to obfuscate the dangerous behaviour during
tests.

If stronger forms of invariant are required, e.g. to take into account interruptions, specific
modelisation choices or dedicated techniques are to be used (cf.
\cite{DBLP:conf/fm/AndronickCP05}).

\subsection{About partial specifications}\label{partial_spec}

Another aspect of a formal specification of a secure system to check is \emph{totality}: is
the behaviour of the system specified in any possible circumstance? It is frequent in formal
methods to define partial specifications -- either to represent a form of contract (a
condition to be realised before having the right to use the system) or a form of freedom left
to the developer (because the systems is not planned to be used in such conditions or because
the result is irrelevant). If the first interpretation can be considered during formal
developments, the second one becomes the only relevant one once leaving the abstract world of
formal methods to tackle with implemented systems. And the extent of the freedom given to the
developer is easily underestimated, as illustrated in the following examples.

We start by two specifications of the {\small$\fid{head}$} function (returning the first
element of a list of natural values) in \emph{Coq}, in the strong specification
style\footnote{In which the return value of a function is described as satisfying a property.}:
\[\small\begin{array}{l}
\fid{head}_1(l\!:\!\fid{list}\:\NAT)(p\!:\!l\!\not=\![\:])\!:\!
 \{x\!:\!\NAT\:|\:\exists\:l'\!:\!\fid{list}\:\NAT,\:l\!=\!x\!::\!l'\}.\\
\fid{head}_2(l\!:\!\fid{list}\:\NAT)\!:\!
 \{x\!:\!\NAT\:|\:l\!\not=\![\:]\to\exists\:l'\!:\!\fid{list}\:\NAT,\:l\!=\!x\!::\!l'\}.\\
\end{array}\]
Both specifications ensure that the function, called upon a non empty list, will return the
head element. Yet the first specification is associated to a precondition, the parameter
{\small$p$} being a proof that the list parameter {\small$l$} is  not empty -- making it
\emph{impossible} to call {\small$\fid{head}_1$} over an empty list as it would not be
possible to build such a proof. The second specification is on the contrary partial, allowing
to use {\small$\fid{head}_2$} with an empty list but not constraining the result in such a
case (except for being a natural value).

The point is that these two specifications are not so different: all the logical parts of a
\emph{Coq} development are eliminated at \emph{extraction} (the process that extract proved
programs). This is not specific to \emph{Coq}: by nature, logical contents in a formal
development are not computable and have therefore to be discarded in some way before being
able to produce a program. And it is easy to implement both specifications in a way that
produces the same following \emph{OCaml} code, where {\small$\fid{secret}$} is any value the
malicious developer would care to export:
\[\small\begin{array}{l}
\Cmd{let}\:\fid{head}=\Cmd{function}\:[\:]\to\fid{secret}\:|\:h::\_\to h
\end{array}\]

We illustrate the same concern in \emph{B} by the specification of a file system manager. We
define the sets {\small$\fid{USR}$} (users), {\small$\fid{Fil}\!\subseteq\!\fid{FIL}$}
(files), {\small$\fid{CNT}$} (contents) and {\small$\fid{RGT}$} (access rights).
{\small$\fid{Cnt}$} associates for any file a content, {\small$\fid{Rgt}$} associates for a
user and a file the rights, and {\small$\fid{cpt}$} gives the number of existing files.
Various operations to create, delete or access the files are assumed to be specified but are
not detailed here, except for {\small$\fid{read}$}:
\[\small\begin{array}{l}
\Cmd{MACHINE}\:\fid{filesystem}\\
\Cmd{SETS}\:\fid{USR};\fid{FIL};\fid{CNT};\fid{RGT}\!=\!\{r,w\}\\
\Cmd{CONSTANTS}\:\fid{cnul}\\
\Cmd{PROPERTIES}\:\fid{cnul}\!\in\!\fid{CNT}\\
\Cmd{VARIABLES}\:\fid{Fil},\fid{Cnt},\fid{Rgt},\fid{cpt}\\
\Cmd{INVARIANT}\:\fid{Fil}\!\subseteq\!\fid{FIL}\land\\
\quad\quad\quad\quad\quad\quad\quad\:
\fid{Cnt}\!\in\!\fid{Fil}\to\fid{CNT}\land\\
\quad\quad\quad\quad\quad\quad\quad\:
\fid{Rgt}\!\subseteq\!(\fid{USR}\!\times\!\fid{Fil})\!\times\!\fid{RGT}\land\\
\quad\quad\quad\quad\quad\quad\quad\:
\fid{cpt}\!=\!\Cmd{card}(\fid{Fil})\\
\Cmd{INITIALISATION}\:\fid{Fil}\!:=\!\emptyset\parallel \fid{Cnt}\!:=\!\emptyset\parallel
                      \fid{Rgt}\!:=\!\emptyset\parallel \fid{cpt}\!:=\!0\\
\Cmd{OPERATIONS}\\
\quad\ldots\\
\quad \fid{out}\gets \fid{read}(f,u)\triangleq\\
\quad\quad\Cmd{PRE}\:f\!\in\!\fid{Fil}\land u\!\in\!\fid{USR}\:\Cmd{THEN}\\
\quad\quad\quad\Cmd{IF}\:((u\!\mapsto\!f)\!\mapsto\!r)\!\in\!\fid{Rgt}\:\Cmd{THEN}\:
 \fid{out}\!:=\!\fid{Cnt}(f)\\
\quad\quad\quad\quad\quad\quad\quad\quad\quad\quad\quad\quad\quad
\Cmd{ELSE}\:\fid{out}\!:=\!\fid{cnul}\\
\quad\ldots\\
\end{array}\]
{\small$\fid{read}$} is specified as returning the content of a file {\small$f$}, provided
that the user {\small$u$} has the right to read it. Yet it is only partially specified, as we
do not describe what happens when the file does not exist. Any call of {\small$\fid{read}$}
implemented in \emph{B} would be associated to a proof obligation to ensure that the
precondition is met, but this constraint goes as far as goes the use of the \emph{B}. So let's
assume the following malicious refinement of {\small$\fid{read}$} is called over a non
existing file:
\[\small\begin{array}{l}
\fid{out}\gets \fid{read}(f,u)\triangleq\\
\quad\Cmd{IF}\:f\!\in\!\fid{Fil}\:\Cmd{THEN}\\
\quad\quad\Cmd{IF}\:((u\!\mapsto\!f)\!\mapsto\!r)\!\in\!\fid{Rgt}\:
\Cmd{THEN}\:\fid{out}\!:=\!\fid{Cnt}(f)\\
\quad\quad\quad\quad\quad\quad\quad\quad\quad\quad\quad\quad\!
\Cmd{ELSE}\:\fid{out}\!:=\!\fid{cnul}\\
\quad\Cmd{ELSE}\:\fid{Fil}\!:=\!\fid{Fil}\!\cup\!\{f_{S}\}\parallel\\
\quad\quad\quad\quad\:\:\!\!
\fid{Cnt}\!:=\!\fid{Cnt}\!\cup\!\{f_{S}\!\mapsto\!S\}\parallel\\
\quad\quad\quad\quad\:\:\!\!
\fid{Rgt}\!:=\!\fid{Rgt}\!\cup\!\{(\fid{eni}\!\mapsto\!f_{S})\!\mapsto\!r\}\\
\end{array}\]
Whereas the specification of {\small$\fid{read}$} was \emph{apparently} passive (not modifying
the state), this refinement creates a file {\small$f_S$} storing a (confidential) value
{\small$S$}, file only accessible by an arbitrary user {\small$\fid{eni}$} invented by the
developer. Furthermore the invariant is broken as {\small$f_S$} is created yet not accounted
for in {\small$cpt$}, that is {\small$f_S$} is virtually invisible for the system. Note also
that defining the returned value when the file does not exist is not even required by
\emph{B}; a malicious developer may however prefer to return {\small$\fid{cnul}$} for a
better obfuscation of its code.

Clearly, a partial specification cannot enforce security, and one should favor a total (and
\emph{defensive}) specification. In \emph{B} this would translate into using a
{\small$\Cmd{IF}$} instead of a {\small$\Cmd{PRE}$}. When the condition associated to an
{\small$\Cmd{IF}$} substitution is not satisfied, the {\small$\Cmd{ELSE}$} branch is executed
-- if it is absent it is equivalent to a {\small$\Cmd{skip}$} substitution, that is it
\emph{enforces} to do nothing. On the contrary when the condition associated to a
{\small$\Cmd{PRE}$} substitution is not satisfied, there is \emph{absolutely no guarantee}
about the result. Note that the defensive approach (with redundant checks) is an
implementation of the \emph{defence in depth} concept.

\subsection{About elusive properties}\label{elusive_prop}

For our next point, we would like to emphasise that some concepts often encountered in
security can be difficult to express in a formal specification. Confidentiality is a good
example: while a formal specification may appear to \emph{implicitly} provide confidentiality,
one should be extremely careful about its exact meaning, as illustrated by the following
example of a login manager in \emph{B}.

The system state is defined by {\small$\fid{Acc}\!\subseteq\!\fid{USR}$} the accounts,
{\small$\fid{log}$} to identify the currently logged account ({\small$\fid{nouser}$} encoding
no opened session), and {\small$\fid{Pwd}$} to associate to any account a password. This last
piece of information is confidential and should not be disclosed. Operations (not detailed in
this paper) allows to log, exit, create or destroy an account, with only the log operation
specified as depending upon {\small$\fid{Pwd}$} to represent the confidentiality of this data.
The operation {\small$\fid{accounts}$}, detailed here, returns the existing accounts:
\[\small\begin{array}{l}
\Cmd{MACHINE}\:\fid{login}\\
\Cmd{SETS}\:\fid{USR};\fid{PWD}\\
\Cmd{CONSTANTS}\:\fid{root},\fid{nouser}\\
\Cmd{PROPERTIES}\:\fid{root}\!\in\!\fid{USR}\land
                  \fid{nouser}\!\in\!\fid{USR}\!\setminus\!\{\fid{root}\}\\
\Cmd{VARIABLES}\:\fid{Acc},\fid{log},\fid{Pwd}\\
\Cmd{INVARIANT}\\
\quad \fid{Acc}\!\subseteq\!\fid{USR}\land\fid{root}\!\in\!\fid{Acc}\land
      \fid{nouser}\!\not\in\!\fid{Acc}\land\\
\quad \fid{log}\!\in\!\fid{Acc}\!\cup\!\{\fid{nouser}\}\land
      \fid{Pwd}\!\in\!\fid{Acc}\!\to\!\fid{PWD}\\
\Cmd{INITIALISATION}\\
\quad\fid{Acc}\!:=\!\{\fid{root}\}\parallel\fid{log}\!:=\!\fid{nouser}\parallel
     \fid{Pwd}\!:\in\!\{\fid{root}\}\!\to\!\fid{PWD}\\
\Cmd{OPERATIONS}\\
\quad\ldots\\
\quad \fid{out}\gets\fid{accounts}\triangleq\\
\quad\quad\Cmd{IF}\:\fid{log}\!\in\!\fid{Acc}\:\Cmd{THEN}\\
\quad\quad\quad\Cmd{ANY}\:s\:\Cmd{WHERE}\:s\!\in\!\Cmd{seq}(\fid{USR})\land\\
\quad\quad\quad\quad\quad\quad\quad\quad\quad\quad\quad\:\:
                                          \Cmd{ran}(s)\!=\!\fid{Acc}\land\\
\quad\quad\quad\quad\quad\quad\quad\quad\quad\quad\quad\:\:
\Cmd{size}(s)\!=\!\Cmd{card}(\fid{Acc})\\
\quad\quad\quad\Cmd{THEN}\:\fid{out}\!:=\!s\\
\quad\quad\Cmd{ELSE}\:\fid{out}\!:=\!\emptyset\\
\quad\ldots\\
\end{array}\]
Input and output values being not refinable in \emph{B} (cf. Par. \ref{B}), the type of the
return value of {\small$\fid{accounts}$} has to be finalised in the specification. In our
example, we have chosen to implement the set {\small$\fid{Acc}$} returned by
{\small$\fid{accounts}$} as a list (or sequence in the \emph{B} terminology) {\small$s$} of
values of {\small$\fid{USR}$}; {\small$\Cmd{ran}(s)\!=\!\fid{Acc}$} ensures that the same
values appear in {\small$\fid{Acc}$} and {\small$s$},
{\small$\Cmd{size}(s)\!=\!\Cmd{card}(\fid{Acc})$} that the length of the list
{\small$s$} is equal to the cardinal of {\small$\fid{Acc}$}. The proposed malicious refinement
of {\small$\fid{accounts}$} is the following one:
\[\small\begin{array}{l}
\quad \fid{out}\gets\fid{accounts}\triangleq\\
\quad\quad\Cmd{IF}\:\fid{log}\!\in\!\fid{Acc}\:\Cmd{THEN}\\
\quad\quad\quad\Cmd{ANY}\:s\:\Cmd{WHERE}\:s\!\in\!\Cmd{seq}(\fid{USR})\land\\
\quad\quad\quad\quad\quad\quad\quad\quad\quad\quad\quad\:\:
                                          \Cmd{ran}(s)\!=\!\fid{Acc}\land\\
\quad\quad\quad\quad\quad\quad\quad\quad\quad\quad\quad\:\:
                                          \Cmd{size}(s)\!=\!\Cmd{card}(\fid{Acc})\\
\quad\quad\quad\Cmd{THEN}\:\Cmd{IF}\:\fid{Pwd}(\fid{root})\!<\!\fid{guess}\\
\quad\quad\quad\quad\quad\quad\:\:\:\!
\Cmd{THEN}\:\fid{out}\!:=\!\Cmd{sort}(s)\\
\quad\quad\quad\quad\quad\quad\:\:\:\!
\Cmd{ELSE}\:\fid{out}\!:=\!\Cmd{rev}(\Cmd{sort}(s))\\
\quad\quad\Cmd{ELSE}\:\fid{out}\!:=\!\emptyset\\
\end{array}\]
where {\small$\fid{guess}$} is a new variable controlled by the malicious developer.
Combining calls to {\small$\fid{accounts}$} and changes of {\small$\fid{guess}$}, one can
quickly derive {\small$\fid{Pwd}(\fid{root})$} through the artificial dependency introduced in
the returned value.

This example illustrates a \emph{covert channel} exploit \cite{DBLP:journals/cacm/Lampson73},
as discussed in \cite{breakingmodel}. Even if the implementation stores {\small$\fid{Pwd}$} in
a private memory location protected by a trusted operating system -- a rather optimistic
assumption -- its confidentiality cannot be guaranteed without a form of control over
dependencies (e.g. considering \emph{data-flow}).

It is of course possible to impose a \emph{complete} (or \emph{monomorphic}) specification
\cite{schonegge-proof} -- a deterministic specification, enforcing the extensional behaviour
of any implementation. A complete specification would not let any freedom to the developer
and thus would ensure that there is no \emph{covert channel} to be exploited. In our example,
a complete specification would for example require {\small$s$} to be sorted in ascending
order. This is however an impractical technical solution, an indirect mean to ensure
confidentiality. Furthermore completeness is not expressible in the \emph{B} specification
language (or in most languages considered in this paper), is generally undecidable and is
\emph{not} stable by refinement of the representation of the data -- e.g. refining a set by an
ordered structure.

It is also possible to better control dependencies in \emph{B} by specifying operations using
constant functions. The following modified specification claims that the operation
{\small$\fid{accounts}$} behaves like a function depending only upon the set
{\small$\fid{Acc}$} and returning a list of values of {\small$\fid{USR}$}:
\[\small\begin{array}{l}
\Cmd{CONSTANTS}\:\ldots,\fid{fct}\\
\Cmd{PROPERTIES}\:
 \ldots\land\fid{fct}\!\in\!\mathbb{P}(\fid{USR})\!\to\!\Cmd{seq}(\fid{USR})\\
\Cmd{OPERATIONS}\\
\quad \fid{out}\gets\fid{accounts}\triangleq\\
\quad\quad\Cmd{IF}\:\fid{log}\!\in\!\fid{Acc}\:\Cmd{THEN}\:\fid{out}\!:=\!\fid{fct}(\fid{Acc})
 \:\Cmd{ELSE}\:\fid{out}\!:=\!\emptyset\\
\end{array}\]
This approach is not yet fully satisfactory as only the dependencies for the \emph{result} are
described (the extensional point of view). It is therefore still possible to affect the
\emph{behaviour} of {\small$\fid{accounts}$}, as in this valid refinement:
\[\small\begin{array}{l}
\fid{out}\gets\fid{accounts}\triangleq\\
\quad\fid{out}:=\fid{encode}(\fid{Pwd}(\fid{root}));\\
\quad \Cmd{IF}\:\fid{Pwd}(\fid{root})\!<\!\fid{guess}
 \:\Cmd{THEN}\:\fid{wait}(10)\:\Cmd{ELSE}\:\fid{wait}(20);\\
\quad\Cmd{IF}\:\fid{log}\!\in\!\fid{Acc}\:\Cmd{THEN}\:\fid{out}\!:=\!\fid{fct}(\fid{Acc})
 \:\Cmd{ELSE}\:\fid{out}\!:=\!\emptyset\\
\end{array}\]
In this refinement the malicious developer implements both a \emph{timed channel} as well as
a possibly observable transient state of the output.

This illustration is just intended to show why, in some cases, expressing confidentiality
can be difficult. For such properties, complementary approaches should be considered, based
e.g. on dependency calculus or non-interference \cite{abadi99core,goguen82}, and associated to
standard code analysis. Note that confidentiality is often formally addressed through access
control enforced by a form of \emph{monitor}, that is according to the \emph{Orange Book} a
tamperproof, unavoidable, and `simple enough to be trusted' mechanism filtering accesses (cf.
recent discussions in
\cite{DBLP:conf/b/BenaissaCM07,DBLP:conf/b/Haddad07,DBLP:conf/b/HoffmannHGB07}). Such a
monitor can itself implement this type of \emph{covert channel} attacks if it is poorly
specified. Note also that the confidence in a system implementing a monitor relies on the
confidence in the information used by this monitor, such as the source of an access request
(that would require a form of \emph{authentication}) as well as the level of protection
required by the accessed object (a meta-information whose origin is generally unclear, but for
which effective implementations such as \emph{security labels} protected in integrity have
been proposed).

We mention authentication and integrity to point out another source of rather elusive
properties, that is the characterisation of cryptographic functions. For example, a
(cryptographic) hash function {\small$H$} is such that:
\begin{itemize}
\item given {\small$h$} it is not possible to find {\small$x$} s.t. {\small$H(x)\!=\!h$};
\item given {\small$x$} it is not possible to find {\small$y\!\not=\!x$} s.t.
{\small$H(x)\!=\!H(y)$};
\item it is not possible to find {\small$x\!\not=\!y$} s.t. {\small$H(x)\!=\!H(y)$}.
\end{itemize}
The first property, for example, guarantees the security of the \emph{Unix} login scheme;
being able to specify a hash function (without giving any details on its implementation) by
formally describing these properties has therefore some interest to certify such a scheme. Yet
these properties appear to be rather \emph{difficult} to express formally. A naive translation
of the last property would just say that {\small$H$} is injective, which is false (as
{\small$H$} projects an infinite set in a finite set of binary words of fixed length) and
would lead to an inconsistent specification. Formally expressing such properties is possible,
but generally less straightforward than one may expect.

\subsection{About the refinement paradox}\label{refinement_paradox}

Most of the examples detailed in Pars. \ref{partial_spec} and \ref{elusive_prop} are
illustrations of what is often referred to as the \emph{refinement paradox}: some properties
are preserved by refinement (safety ones generally are), other are not (security ones).

Back to the discussion of Par. \ref{elusive_prop}, the most simple example of `devious'
refinement that we can exhibit in \emph{B} is the following one:
\[\small\begin{array}{l}
\Cmd{MACHINE}\:\fid{Boolean}\\
\Cmd{OPERATIONS}\:\fid{out}\!\gets\!\fid{go}\triangleq
\fid{out}\!:=\!\fid{true}\talloblong\fid{out}\!:=\!\fid{false}
\end{array}\]
This machine is a very simple one, having no state and defining a single operation
{\small$\fid{go}$} returning a boolean value. There are of course two straightforward
refinements:
\[\small\begin{array}{l}
\Cmd{MACHINE}\:\fid{Boolean\_True}\\
\Cmd{REFINES}\:\fid{Boolean}\\
\Cmd{OPERATIONS}\:\fid{out}\!\gets\!\fid{go}\triangleq\fid{out}\!:=\!\fid{true}
\end{array}\]
\[\small\begin{array}{l}
\Cmd{MACHINE}\:\fid{Boolean\_False}\\
\Cmd{REFINES}\:\fid{Boolean}\\
\Cmd{OPERATIONS}\:\fid{out}\!\gets\!\fid{go}\triangleq\fid{out}\!:=\!\fid{false}
\end{array}\]
Yet it also accepts other refinements, such as the following one:
\[\small\begin{array}{l}
\Cmd{MACHINE}\:\fid{Boolean\_Covert\_Channel}\\
\Cmd{REFINES}\:\fid{Boolean}\\
\Cmd{VARIABLE}\:\fid{dump}\\
\Cmd{INVARIANT}\:\fid{dump}\!\in\!\NAT\\
\Cmd{INITIALISATION}\:\fid{dump}\!:=\!\fid{private\_key}\\
\Cmd{OPERATIONS}\:\fid{out}\!\gets\!\fid{go}\triangleq
\Cmd{IF}\:\fid{dump}\:\Cmd{mod}\:2\!=\!0\\
\quad\quad\quad\quad\quad\quad\quad\quad\quad\quad\quad\quad\quad
\Cmd{THEN}\:\fid{out}\!:=\!true\\
\quad\quad\quad\quad\quad\quad\quad\quad\quad\quad\quad\quad\quad
\Cmd{ELSE}\:\fid{out}\!:=\!false;\\
\quad\quad\quad\quad\quad\quad\quad\quad\quad\quad\quad\quad\quad
\fid{dump}\!:=\!\fid{dump}/2
\end{array}\]
One should not believe that the refinement paradox is specific to those methods which are
providing an explicit form of refinement, such as \emph{B} or \emph{Z} for example. Our
devious refinements include implicitly a non functional refinement of the representation of
data: we accept several implementations as representing a single abstract value of the
specification. This intuitively describes why some variables are \emph{hidden} at the
specification level. From this intuition, we suggest the following counterpart in \emph{Coq}
of the refinement paradox. Let's consider the example of the specification of booleans as an
\emph{Abstract Data Type}, with the equality and a boolean function:
\[\small\begin{array}{l}
\Cmd{Module\:Type}\:\fid{Boolean\_Function}.\\
\quad\Cmd{Parameter}\:B\!:\!\Cmd{Set}.\\
\quad\Cmd{Parameters}\:\top\:\bot\!:\!B.\\
\quad\Cmd{Parameter}\:\equiv:\!B\!\to\!B\!\to\!\Cmd{Prop}.\\
\quad\Cmd{Hypothesis}\:\fid{refl}\!:\!\forall\:(b\!:\!B),\:b\!\equiv\!b.\\
\quad\Cmd{Hypothesis}\:\fid{sym}\!:\!
 \forall\:(b_1\:b_2\!:\!B),\:b_1\!\equiv\!b_2\!\to\!b_2\!\equiv\!b_1.\\
\quad\Cmd{Hypothesis}\:\fid{tran}\!:\!
 \forall\:(b_1\:b_2\:b_3\!:\!B),
 \:b_1\!\equiv\!b_2\!\to\!b_2\!\equiv\!b_3\!\to\!b_1\!\equiv\!b_3.\\
\quad\Cmd{Hypothesis}\:\fid{inj}\!:\!\lnot\top\!\equiv\!\bot.\\
\quad\Cmd{Hypothesis}\:\fid{surj}\!:\!\forall\:(b\!:\!B),\:
 b\!\equiv\!\top\lor b\!\equiv\!\bot.\\
\quad\Cmd{Parameter}\:\fid{fnc}\!:\!B\!\to\!B.\\
\Cmd{End}\:\fid{Boolean\_Function}.
\end{array}\]
The straightforward refinement of this specification is of course to implement {\small$B$} as
{\small$\BOOL$}, the \emph{Coq} type of booleans, and to implement {\small$\fid{fnc}$} as one
of the four possible boolean functions ({\small$\fid{true}$}, {\small$\fid{false}$},
{\small$\fid{identity}$} or {\small$\fid{not}$}). But a devious implementation gives much more
freedom; we can for example choose to implement {\small$B$} as {\small$\NAT$}, even values
representing {\small$\bot$} and odd values representing {\small$\top$}:
\[\small\begin{array}{l}
\Cmd{Module}\:\fid{Covert\_Channel}\!:\!\fid{Boolean\_Function}.\\
\quad\Cmd{Definition}\:B\!:=\!\NAT.\\
\quad\Cmd{Definition}\:\bot\!:=\!0.\:\Cmd{Definition}\:\top\!:=\!1.\\
\quad\Cmd{Definition}\:\equiv\!(b_1\:b_2:B)\!:=\!(b_1\!+\!b_2\:\Cmd{mod}\:2\!=\!0).\\
\quad\ldots\\
\quad\Cmd{Definition}\:\fid{fnc}(b\!:\!B)\!:\!B\!:=\!
\Cmd{match}\:((b/2)\:\Cmd{mod}\:4)\:\Cmd{with}\\
\quad\quad|\:0\:\Rightarrow\:\bot\\
\quad\quad|\:1\:\Rightarrow\:\top\\
\quad\quad|\:2\:\Rightarrow\:b\\
\quad\quad|\:\_\:\Rightarrow\:b\!+\!1\\
\quad\quad\Cmd{end}.\\
\Cmd{End}\:\fid{Covert\_Channel}.
\end{array}\]
This implementation introduces a new dimension in the representation of the data, which is
hidden at specification level and can be used by a malicious developer to store information
and modify results: {\small$\fid{fnc}$} now emulates any of the boolean functions.

Note that the term of refinement paradox may be considered an overstatement, provided the
presentation of refinement in Par. \ref{refinement}. Clearly the very concept of refinement is
extensional, whereas on the contrary confidentiality can be considered as intensional: rather
than describing \emph{what} a result should be, it aims at constraining \emph{how} a result is
produced (in this case, without depending upon the confidential value). Similarly, if
refinement is intended to preserve properties described in a specification, it does not aim at
preserving properties of the specification itself, or any other form of
\emph{meta-properties}; so the fact that for example completeness is not preserved should not
be a surprise.

\section{Building on Sand?}\label{building_sand}

In Par. \ref{invalid_spec}, we have shown possible consequences of inconsistent
specifications. Obviously similar or worse consequences can result from other sources of
inconsistencies, such as a bug in the tool implementing the formal method, or a mistake in the
theory of the formal method itself. For a malicious developer, a paradox (a flaw in the logic
that can be used to prove at the same time both {\small$P$} and {\small$\lnot P$}) discovered
in a theory or in a tool can be used to prove any property about any development, that is to
implement any unpleasant behaviour while getting a certification.

When trying to assess the level of confidence one may have in the result of a formal
development, the question of the validity of the tool and of the theory should therefore be
addressed.

\subsection{About the logic}\label{logic_error}

In \cite{DBLP:conf/lpar/JaegerD07} a deep embedding (cf. \cite{gor:2,azu:1}) of the \emph{B}
logic in \emph{Coq} is described, that is intuitively a form of \emph{B}
\emph{virtual machine} developed in \emph{Coq} with the objective to check the validity of
the \emph{B} logic. While this deep embedding has not identified any paradox\footnote{The
consistency of the \emph{B} logic has not been proved either.}, it has shown that the
following `theorems' from \cite{abr:1} are in fact not provable using the defined logic:
\[\small\begin{array}{l}
E_1\!\!\mapsto\!\!F_1\!=\!E_2\!\!\mapsto\!\!F_2\Rightarrow E_1\!=\!E_2\\
E_1\!\!\mapsto\!\!F_1\!=\!E_2\!\!\mapsto\!\!F_2\Rightarrow F_1\!=\!F_2\\
S_1\!\subseteq\!S_2 \land T_1\!\subseteq\!T_2 \Rightarrow
S_1\!\times\!T_1\!\subseteq\!S_2\!\times\!T_2
\end{array}\]
These results are not provable because of the definition of the \emph{B} inference rules,
which are not sufficiently precise regarding the formal definition of what is a cartesian
product. To our knowledge, the fact that these results were not valid in \emph{B} was not
known by the \emph{B} community. Being apparently trivial, they were never checked and have
been integrated for example in provers for the \emph{B} logic. That means, at a fundamental
level, that these results were in fact taken as additional axioms, without people knowing it
-- an approach that could have created a paradox in the logic.

Further investigations have emphasised another form of subtile glitch that may appear in the
theory of a formal method. As pointed out  in Par. \ref{formal_spec}, formal methods allow for
multiple descriptions of a system as well as the verification of the similarity of these
descriptions. This is sometimes obtained by defining several semantics for a single construct.

In \emph{B}, substitutions of the \emph{GSL} (used to write operations) are defined as
\emph{predicate transformers}, that is a logical semantic. On the other hand the substitutions
of the \emph{B0} sub-language are used for implementation and also have an operational
semantic. This is the case of the
{\small$\Cmd{WHILE}\:P\:\Cmd{DO}\:S\:\Cmd{INVARIANT}\:I\:\Cmd{VARIANT}\:V$} substitution,
illustrated in \cite{abr:1} by the extraction of the minimum of a non-empty set of natural
values:
\[\small\begin{array}{l}
x\!:=\!0;\begin{array}{l}\Cmd{WHILE}\:x\!\not\in\!S\\
                           \quad\Cmd{DO}\:x\!:=\!x\!+\!1\\
                           \quad\Cmd{INVARIANT}\:x\!\in\![0,\Cmd{min}(S)]\\
                           \quad\Cmd{VARIANT}\:\Cmd{min}(S)\!-\!x\\
                           \Cmd{END}\end{array}
\end{array}\]
Using the definition of the {\small$\Cmd{WHILE}$} substitution as a predicate transformer, one
can indeed show that this substitution realises (that is, transforms into a tautology) the
predicate {\small$x\!=\!\Cmd{min}(S)$}. In other words the substitution is proven to extract
the minimum in any case of use (provided {\small$S\!\not=\!\emptyset$}).

By denoting {\small$\llbracket\rrbracket$} the translation producing a \emph{C} program from a
\emph{B0} substitution, the operational semantic is defined by:
\[\begin{array}{l}
\left\llbracket
 \begin{array}{l}
 \Cmd{WHILE}\:P\\
 \Cmd{DO}\:S\\
 \Cmd{INVARIANT}\:I\\
 \Cmd{VARIANT}\:V
 \end{array}
\right\rrbracket=\Cmd{while}\:\llbracket P\rrbracket\{\llbracket S\rrbracket\}
\end{array}\]
The interesting point is that this semantic forgets {\small$I$} (the loop invariant) and
{\small$V$} (the loop variant) that are pure logical contents, important for the proofs (e.g.
of termination) but irrelevant for the execution.

Modifying the invariant does not change the program (the operational semantic) and should
therefore only have limited impact on the logical semantic. The surprise is that by replacing
in the previous example the invariant {\small$m\!\in\![0,\Cmd{min}(S)]$} by
{\small$m\!\in\!\NAT$}, less precise but still correct, the logical semantic is
\emph{radically} modified. This modified logical semantic leads to a refutation of the
previous proposition, that is it indicates that the substitution is not always extracting the
minimum. A rather strange conclusion, as both versions of the logical semantic describe the
same program.

We have also identified a similar concern with \emph{Coq}. In this case there is a single
language, mixing logical and computational constructs, an extraction mechanism allowing for
the elimination of the former to derive from the latter a program in a functional language,
e.g. in \emph{OCaml}.

As already pointed out in Par. \ref{invalid_spec}, an inductive definition such as
{\small$\Cmd{Inductive}\:E\!:\!\Cmd{Set}\!:=\!\fid{nxt}\!:\!E\!\to\!E$} lacks an atomic
constructor and is therefore empty. Emptyness is not, by itself, inconsistent but makes
possible to prove any result of the form {\small$\forall\:(e\!:\!E),\:P$}. Its extraction in
\emph{OCaml} is a straightforward translation to
{\small$\Cmd{type}\:E\!=\!\fid{Nxt}\:\Cmd{of}\:E$}. The interesting point is that this
\emph{OCaml} type is \emph{not} empty, as it contains the value
{\small$\Cmd{let}\:\Cmd{rec}\:e\!=\!\fid{Nxt}(e)$}, not valid in \emph{Coq} but making
possible to use a program extracted from a fully certified \emph{Coq} library with unexpected
(and therefore unwanted) behaviours.

It is beyond the scope of this paper to further discuss these questions, once noted that any
such bias is a potential weakness usable by a malicious developer (or a trap for an honest
but inattentive developer). These remarks are not intended to criticize the tremendous work
represented by the full development of the theories supporting formal methods. They however
justify the interest in mechanically checking such theories, pursuing works described e.g. in
\cite{cha:1,bod:1b,Barras99}.

\subsection{About the tools}\label{tool_error}

Beyond the concerns about the theory, one may also question the validity of the tool
implementing a formal method. For example a prover can be incomplete (unable to prove results
valid in the theory) or incorrect (able to prove results unprovable in the theory), the latter
being more worrying, at least from the evaluation and certification perspective, as it may
lead to an artificial paradox. And indeed such paradoxes have been discovered in well
established tools.

Clearly, implementing a formal method is a difficult task, dealing not only with completeness,
correctness, but also with performance, automation, and ergonomy. In our view, the (potential)
existence of bugs in a tool does not mean that it should not be used, but that the provided
results should be considered with some care, and possibly verified by other mechanisms. This
is addressed for example by \cite{DBLP:conf/lpar/JaegerD07,rid:1}.

\section{Stepping Out of the Model}\label{stepping_out}

We have discussed at length some concerns regarding the formal development of secure systems,
through questionning paradoxes in the theory, bugs in the tools or more simply by identifying
gotchas in the specifications. Let's now assume that we have been able to produce a consistent
specification with security properties correctly expressed, and a compliant implementation
whose all proof obligations have been discharged, using a well-established formal method and a
trusted tool -- that is, we finally have a \emph{proven security} system. That does \emph{not}
mean however that the system is secure, but that any attack has to contradict at least one of
the hypotheses (a good heuristic for those willing to attack formally validated systems).

Preconditions, for example, are hypotheses whose violation can be devastating, as illustrated
in Par. \ref{partial_spec}. But one should take care also to identify all the \emph{implicit}
hypotheses when developing a system or evaluating its security. Such implicit hypotheses are
not only those that are introduced by the formal method (cf. Par. \ref{invalid_spec}), but
also those that are related to the modelisation choices themselves.

\subsection{About Closure}\label{closure}

A frequent implicit hypothesis is related to the use of closure proofs. For example, proving a
\emph{B} machine requires proving the preservation of its invariant by any of its operations.
This is justified if there is no other way to influence the system state than the provided
operations. The extent to which this is enforced in the real system has to be carefully
analysed. Threats considered during security analysis may reflect actions that are not in the
model (data stored in files by proven applications can be modified by other applications,
signals in electronic circuits can be jammed by \emph{fault injection}, etc). There is no
silver bullet to address this problem; current approaches include defensive style programming,
redundancy, and dysfunctional considerations (e.g. by modelling errors such as unexpected
values or inconsistent states).

\subsection{About Typing}\label{typing}

A second example of implicit hypothesis, much less obvious, is related to types. An adequate
use of types in a specification (for example modelling \emph{IP} addresses and ports as
values of abstract sets rather than natural values) ensures that some forms of error will be
automatically detected (such as using a port where an address is expected). But it is also
important to understand how strong an hypothesis it is, and how easily it can be violated.
Indeed, types are again logical information that have generally no concrete implementation; in
most programming languages, they just disappear at compilation. So, while ill typed operation
calls \emph{cannot} be considered during formal analysis, they are in some cases
\emph{executable}.

A typical example is provided in \cite{DBLP:conf/ches/Clulow03}, describing a flaw in the
\emph{PKCS\#11} API for cryptographic resources, summarised here. A central authority (e.g. a
bank) distributes cryptographic resources to customers. Such a resource can perform
cryptographic operations, {\small$C\!\gets\!\fid{cipher}(M,K)$} to cipher the message
{\small$M$} with the key numbered {\small$K$}, or {\small$M\!\gets\!\fid{uncipher}(C,K)$} for
the inverse operation. The resource never discloses keys to the customer, but permits exchange
of keys with other resources through export of \emph{wrapped} (cyphered) keys using
{\small$D\!\gets\!\fid{export}(K,W)$} where {\small$K$} is the number of the exported key and
{\small$W$} the number of the wrapping key, and {\small$\fid{import}(D,W,K)$} for the inverse
operation (that stores internally the unwrapped key under number {\small$K$} without
disclosing it). In a model where cyphertexts and wrapped keys are of different types, one can
prove that no sequence of calls will disclose a sensitive key. Unfortunately the
implementations of cyphertexts and wrapped keys are indistinguishable, and stored keys are not
tagged with their role. It is so possible to disclose a key {\small$K$} with the (ill-typed)
sequence {\small$D\!\gets\!\fid{export}(K,W);M\!\gets\!\fid{uncipher}(D,W)$}.

This demonstrates that it is important to identify implicit hypotheses associated to the use
of types to detect possible consequences of type violations, or to maintain type information
in the implementation to prevent such attacks.

\section{Conclusion}\label{conclusion}

We summarise and discuss difficulties related to the development of secure systems using
formal methods, identifying -- where possible -- proposals for improvement. The concerns
described in this paper were identified during a systematic review of the process of formal
development, investigating possible difficulties.

A quick read of this paper could seem to imply that the reputation of formal methods to
develop correct systems is overestimated. \emph{This is not our message}. We consider that
formal methods are very efficient tools to obtain high level of assurance and confidence for
the development of systems in general, and of secure systems in particular.

Yet to fully benefit from such tools, one has to understand their strengths but also their
limitations. Pretending that proven secure systems are perfectly secure is nothing more than
a renewed version of the first myth about formal methods pointed out in
\cite{10.1109/52.57887}, and is to the least inadequate; in fact, we consider that such a
claim is detrimental to formal methods. Taking this into account, we expect our proposals to
help, where possible, for improving the quality of formal specifications and the adequacy of
formal developments of secure systems (in some cases relying on other methodologies or
technologies); our second expectation is to shed some light on the difficulties to at least
allow for a better evaluation of the genuine level of confidence obtained through the use of
formal methods.

\subsubsection*{Nota}{\small An extended version of this paper is available in French language
at \cite{ssigouv}}.



%


\bibliographystyle{IEEEtran}
\bibliography{Traps}

\begin{thebibliography}{10}
\providecommand{\url}[1]{#1}
\csname url@samestyle\endcsname
\providecommand{\newblock}{\relax}
\providecommand{\bibinfo}[2]{#2}
\providecommand{\BIBentrySTDinterwordspacing}{\spaceskip=0pt\relax}
\providecommand{\BIBentryALTinterwordstretchfactor}{4}
\providecommand{\BIBentryALTinterwordspacing}{\spaceskip=\fontdimen2\font plus
\BIBentryALTinterwordstretchfactor\fontdimen3\font minus
  \fontdimen4\font\relax}
\providecommand{\BIBforeignlanguage}[2]{{%
\expandafter\ifx\csname l@#1\endcsname\relax
\typeout{** WARNING: IEEEtran.bst: No hyphenation pattern has been}%
\typeout{** loaded for the language `#1'. Using the pattern for}%
\typeout{** the default language instead.}%
\else
\language=\csname l@#1\endcsname
\fi
#2}}
\providecommand{\BIBdecl}{\relax}
\BIBdecl

\bibitem{iec61508}
{IEC 61508}, ``Functional safety of electrical, electronic, programmable
  electronic safety-related systems,'' (\url{www.iec.ch/zone/fsafety}).

\bibitem{CommonCriteria}
{ISO/IEC 15408}, ``Common criteria for information technology security
  evaluation,'' (\url{www.commoncriteriaportal.org}).

\bibitem{coq:1}
\BIBentryALTinterwordspacing
{The Coq development team}, \emph{The {Coq} proof assistant reference manual},
  {LogiCal} Project, 2004. [Online]. Available: \url{http://coq.inria.fr}
\BIBentrySTDinterwordspacing

\bibitem{abr:1}
J.~R. Abrial, \emph{The {B-Book} - Assigning Programs to Meanings}.\hskip 1em
  plus 0.5em minus 0.4em\relax Cambridge University Press, Aug. 1996.

\bibitem{focal}
{The {FoCal} development team}, ``The {Focal} project,''
  (\url{focal.inria.fr}).

\bibitem{pvs:1}
\BIBentryALTinterwordspacing
S.~Owre, J.~M. Rushby, , and N.~Shankar, ``{PVS:} {A} prototype verification
  system,'' in \emph{11th International Conference on Automated Deduction
  (CADE)}, ser. Lecture Notes in Artificial Intelligence, D.~Kapur, Ed., vol.
  607.\hskip 1em plus 0.5em minus 0.4em\relax Saratoga, {NY}: Springer-Verlag,
  jun 1992, pp. 748--752. [Online]. Available:
  \url{http://www.csl.sri.com/papers/cade92-pvs/}
\BIBentrySTDinterwordspacing

\bibitem{nip:1}
T.~Nipkow, L.~C. Paulson, and M.~Wenzel, \emph{{Isabelle/HOL} --- A Proof
  Assistant for Higher-Order Logic}, ser. Lecture Notes in Computer
  Science.\hskip 1em plus 0.5em minus 0.4em\relax Springer, 2002, vol. 2283.

\bibitem{beh:1}
P.~Behm, P.~Desforges, and J.~M. Meynadier, ``{M{\'E}T{\'E}OR} : An industrial
  success in formal development.'' in \emph{{B '98}: Proceedings of the Second
  International {B} Conference on Recent Advances in the Development and Use of
  the {B} Method}, ser. Lecture Notes in Computer Science, D.~Bert, Ed., vol.
  1393.\hskip 1em plus 0.5em minus 0.4em\relax Springer, 1998, p.~26.

\bibitem{DBLP:conf/colog/CoquandP88}
T.~Coquand and C.~Paulin, ``Inductively defined types,'' in \emph{Conference on
  Computer Logic}, ser. Lecture Notes in Computer Science, P.~Martin-L{\"o}f
  and G.~Mints, Eds., vol. 417.\hskip 1em plus 0.5em minus 0.4em\relax
  Springer, 1988, pp. 50--66.

\bibitem{mus:1}
L.~Mussat, 2005, private Communication.

\bibitem{CarlierD2008}
M.~Carlier and C.~Dubois, ``Functional testing in the focal environment,'' in
  \emph{Test And Proof (TAP'2008)}, B.~Berckert and R.~Hahnle, Eds., vol.
  4966.\hskip 1em plus 0.5em minus 0.4em\relax LNCS, 2008, pp. 84--98.

\bibitem{B07-JaffuelL}
E.~Jaffuel and B.~Legeard, ``{LEIRIOS} test generator: Automated test
  generation from {B} models.'' in \emph{The 7th International {B} Conference},
  ser. Lecture Notes in Computer Science, J.~Julliand and O.~Kouchnarenko,
  Eds., vol. 4355.\hskip 1em plus 0.5em minus 0.4em\relax Springer, 2007, pp.
  277--280.

\bibitem{ClarkeGL96}
E.~M. Clarke, O.~Grumberg, and D.~E. Long, ``Model checking,'' in \emph{NATO
  ASI DPD}, M.~Broy, Ed., 1996, pp. 305--349.

\bibitem{PlaggeL07}
D.~Plagge and M.~Leuschel, ``Validating {Z} specifications using the
  {ProBAnimator} and model checker,'' in \emph{IFM}, ser. Lecture Notes in
  Computer Science, J.~Davies and J.~Gibbons, Eds., vol. 4591.\hskip 1em plus
  0.5em minus 0.4em\relax Springer, 2007, pp. 480--500.

\bibitem{YuML99}
Y.~Yu, P.~Manolios, and L.~Lamport, ``Model checking {TLA$^{\mbox{+}}$}
  specifications,'' in \emph{CHARME}, ser. Lecture Notes in Computer Science,
  L.~Pierre and T.~Kropf, Eds., vol. 1703.\hskip 1em plus 0.5em minus
  0.4em\relax Springer, 1999, pp. 54--66.

\bibitem{DBLP:conf/lpar/SamerV07}
M.~Samer and H.~Veith, ``On the notion of vacuous truth,'' in \emph{LPAR}, ser.
  Lecture Notes in Computer Science, N.~Dershowitz and A.~Voronkov, Eds., vol.
  4790.\hskip 1em plus 0.5em minus 0.4em\relax Springer, 2007, pp. 2--14.

\bibitem{TFTP04}
C.~Dubois, T.~Hardin, and V.~V. {Donzeau Gouge}, ``{Building certified
  components within FOCAL},'' in \emph{Trends in Functional Programming}, H.-W.
  Loidl, Ed., vol.~5.\hskip 1em plus 0.5em minus 0.4em\relax Bristol, UK:
  Intellect, 2004, pp. 33--48.

\bibitem{TPHOL2003}
D.~Doligez, T.~Hardin, and V.~Pr\'evosto, ``Algebraic structures and dependent
  records,'' in \emph{Proceedings of TPHOL'03}, B.~Rick, Ed., NASA, Hampton,
  USA, 2003.

\bibitem{calc03}
V.~Pr\'evosto and M.~Jaume, ``Making proofs in a hierarchy of mathematical
  structures,'' in \emph{11th Symposium on the Integration of Symbolic
  Computation and Mechanized Reasoning, Calculemus 2003}.\hskip 1em plus 0.5em
  minus 0.4em\relax Aracne, 2003, pp. 89--100.

\bibitem{DBLP:conf/fm/AndronickCP05}
J.~Andronick, B.~Chetali, and C.~Paulin-Mohring, ``Formal verification of
  security properties of smart card embedded source code,'' in \emph{FM}, ser.
  Lecture Notes in Computer Science, J.~Fitzgerald, I.~J. Hayes, and
  A.~Tarlecki, Eds., vol. 3582.\hskip 1em plus 0.5em minus 0.4em\relax
  Springer, 2005, pp. 302--317.

\bibitem{DBLP:journals/cacm/Lampson73}
B.~W. Lampson, ``A note on the confinement problem,'' \emph{Commun. ACM},
  vol.~16, no.~10, pp. 613--615, 1973.

\bibitem{breakingmodel}
\BIBentryALTinterwordspacing
J.~A. Clark, S.~Stepney, and H.~Chivers, ``Breaking the model: Finalisation and
  a taxonomy of security attacks.'' [Online]. Available:
  \url{citeseer.ist.psu.edu/clark04breaking.html}
\BIBentrySTDinterwordspacing

\bibitem{schonegge-proof}
\BIBentryALTinterwordspacing
A.~Sch{\"o}negge, ``Proof obligations for monomorphicity.'' [Online].
  Available: \url{citeseer.ist.psu.edu/27234.html}
\BIBentrySTDinterwordspacing

\bibitem{abadi99core}
\BIBentryALTinterwordspacing
M.~Abadi, A.~Banerjee, N.~Heintze, and J.~G. Riecke, ``A core calculus of
  dependency,'' in \emph{{POPL} '99. Proceedings of the 26th {ACM}
  {SIGPLAN-{SIGACT}} on Principles of programming languages, January 20--22,
  1999, San Antonio, {TX}}, {ACM}, Ed.\hskip 1em plus 0.5em minus 0.4em\relax
  New York, NY, USA: ACM Press, 1999, pp. 147--160. [Online]. Available:
  \url{citeseer.ist.psu.edu/abadi99core.html}
\BIBentrySTDinterwordspacing

\bibitem{goguen82}
J.~Goguen and J.~Meseguer, ``Security policies and security models,'' in
  \emph{IEEE Symposium on Security and Privacy}.\hskip 1em plus 0.5em minus
  0.4em\relax IEEE Press, 1992, pp. 11--20.

\bibitem{DBLP:conf/b/BenaissaCM07}
N.~Bena\"{\i}ssa, D.~Cansell, and D.~M{\'e}ry, ``Integration of security policy
  into system modeling,'' in \emph{B}, ser. Lecture Notes in Computer Science,
  J.~Julliand and O.~Kouchnarenko, Eds., vol. 4355.\hskip 1em plus 0.5em minus
  0.4em\relax Springer, 2007, pp. 232--247.

\bibitem{DBLP:conf/b/Haddad07}
A.~Haddad, ``Meca: A tool for access control models,'' in \emph{B}, ser.
  Lecture Notes in Computer Science, J.~Julliand and O.~Kouchnarenko, Eds.,
  vol. 4355.\hskip 1em plus 0.5em minus 0.4em\relax Springer, 2007, pp.
  281--284.

\bibitem{DBLP:conf/b/HoffmannHGB07}
S.~Hoffmann, G.~Haugou, S.~Gabriele, and L.~Burdy, ``The {B-Method} for the
  construction of microkernel-based systems,'' in \emph{B}, ser. Lecture Notes
  in Computer Science, J.~Julliand and O.~Kouchnarenko, Eds., vol. 4355.\hskip
  1em plus 0.5em minus 0.4em\relax Springer, 2007, pp. 257--259.

\bibitem{DBLP:conf/lpar/JaegerD07}
{\'E}.~Jaeger and C.~Dubois, ``Why would you trust {B} ?'' in \emph{LPAR}, ser.
  Lecture Notes in Computer Science, N.~Dershowitz and A.~Voronkov, Eds., vol.
  4790.\hskip 1em plus 0.5em minus 0.4em\relax Springer, 2007, pp. 288--302.

\bibitem{gor:2}
\BIBentryALTinterwordspacing
{M.J.C. Gordon}, ``Mechanizing programming logics in higher-order logic,'' in
  \emph{Current Trends in Hardware Verification and Automatic Theorem Proving
  ({P}roceedings of the Workshop on Hardware Verification)}, {G.M. Birtwistle}
  and {P.A. Subrahmanyam}, Eds.\hskip 1em plus 0.5em minus 0.4em\relax Banff,
  Canada: Springer-Verlag, Berlin, 1988, pp. 387--439. [Online]. Available:
  \url{citeseer.ist.psu.edu/gordon88mechanizing.html}
\BIBentrySTDinterwordspacing

\bibitem{azu:1}
A.~Azurat and I.~Prasetya, ``A survey on embedding programming logics in a
  theorem prover,'' Institute of Information and Computing Sciences, Utrecht
  University, Tech. Rep. UU-CS-2002-007, 2002.

\bibitem{cha:1}
P.~Chartier, ``Formalisation of {B} in {Isabelle/HOL},'' in \emph{{B '98}:
  Proceedings of the Second International {B} Conference on Recent Advances in
  the Development and Use of the {B} Method}, ser. Lecture Notes in Computer
  Science, D.~Bert, Ed., vol. 1393.\hskip 1em plus 0.5em minus 0.4em\relax
  London, UK: Springer-Verlag, 1998, pp. 66--82.

\bibitem{bod:1b}
J.-P. Bodeveix, M.~Filali, and C.~Mu{\~{n}}oz, ``A formalization of the
  {B-Method} in {Coq} and {PVS},'' in \emph{Electronic Proceedings of the
  B-User Group Meeting at the World Congress on Formal Methods FM 99}, 1999,
  pp. 33--49.

\bibitem{Barras99}
B.~Barras, ``Auto-validation d'un syst{\`e}me de preuves avec familles
  inductives,'' Th{\`e}se de Doctorat, Universit{\'e} Paris~7, Nov. 1999.

\bibitem{rid:1}
T.~Ridge and J.~Margetson, ``A mechanically verified, sound and complete
  theorem prover for first order logic.'' in \emph{TPHOLs}, ser. Lecture Notes
  in Computer Science, J.~Hurd and T.~F. Melham, Eds., vol. 3603.\hskip 1em
  plus 0.5em minus 0.4em\relax Springer, 2005, pp. 294--309.

\bibitem{DBLP:conf/ches/Clulow03}
J.~Clulow, ``On the security of {PKCS\#11},'' in \emph{CHES}, ser. Lecture
  Notes in Computer Science, C.~D. Walter, \c{C}etin Kaya~Ko\c{c}, and C.~Paar,
  Eds., vol. 2779.\hskip 1em plus 0.5em minus 0.4em\relax Springer, 2003, pp.
  411--425.

\bibitem{10.1109/52.57887}
A.~Hall, ``Seven myths of formal methods,'' \emph{IEEE Software}, vol.~07,
  no.~5, pp. 11--19, 1990.

\bibitem{ssigouv}
{DCSSI}, ``Central directorate for information systems security, publications
  page,'' (\url{http://www.ssi.gouv.fr/fr/sciences/publications.html}).

\bibitem{crf:3}
D.~Bert, Ed., \emph{B'98: Recent Advances in the Development and Use of the B
  Method, Second International B Conference, Montpellier, France, April 22-24,
  1998, Proceedings}, ser. Lecture Notes in Computer Science, vol. 1393.\hskip
  1em plus 0.5em minus 0.4em\relax Springer, 1998.

\bibitem{DBLP:conf/lpar/2007}
N.~Dershowitz and A.~Voronkov, Eds., \emph{Logic for Programming, Artificial
  Intelligence, and Reasoning, 14th International Conference, LPAR 2007,
  Yerevan, Armenia, October 15-19, 2007, Proceedings}, ser. Lecture Notes in
  Computer Science, vol. 4790.\hskip 1em plus 0.5em minus 0.4em\relax Springer,
  2007.

\bibitem{DBLP:conf/b/2007}
J.~Julliand and O.~Kouchnarenko, Eds., \emph{B 2007: Formal Specification and
  Development in B, 7th International Conference of B Users, Besan\c{c}on,
  France, January 17-19, 2007, Proceedings}, ser. Lecture Notes in Computer
  Science, vol. 4355.\hskip 1em plus 0.5em minus 0.4em\relax Springer, 2006.

\end{thebibliography}

\end{document}